\documentclass[
reprint,
 amsmath,amssymb,
 aps,
prb,
]{revtex4-2}
\setcitestyle{numbers,square}
\usepackage{amsmath,amssymb}
\usepackage{tabularx}
\usepackage{graphicx}
\usepackage{bmpsize}
\usepackage{bm}
\usepackage{color}
\usepackage{amssymb}
\usepackage[version=3]{mhchem}
\usepackage{newtxmath}
\usepackage{physics}
\usepackage{hyperref}
\usepackage{mathtools}
\usepackage{xcolor}
\usepackage{ulem}
\usepackage{comment}

\begin{document}
\title{Nonequilibrium Dirac condensate in bosonic Kitaev chain}
\author{Kazuki Yamamoto}
\author{Youichi Yanase}
\affiliation{Department of Physics, Kyoto University, Kyoto 606-8502, Japan}
\begin{abstract}
We explore a nonequilibrium Bose-Einstein condensate (BEC) in a bosonic Kitaev chain. 
Within a mean-field framework, 
we show that bosonic superfluid order develops
once the pairing strength exceeds a critical threshold. 
The condensate order parameter adopts an alternating phase between even and odd sites,
giving rise to a spontaneous 
sublattice ordering.
By mapping the emergent sublattice degrees of freedom onto a pseudospin, 
the elementary Bogoliubov excitation is described by a non-Hermitian Dirac Hamiltonian, displaying two prominent features: a re-entrant 
bosonic Andreev bound state localized around boundaries,
and a macroscopic degeneracy of exceptional 
Majorana bosons
at the critical point. 
\end{abstract}
\maketitle


\textbf{\textit{Introduction.}}---
The Kitaev chain~\cite{Kitaev2001}, a one-dimensional lattice model of a spinless $p$-wave superconductor, serves as a foundational paradigm of topological superconductivity~\cite{qi2009time,qi2011topological,sato2017topological,Yamamoto2026_topological} hosting Majorana fermions localized at its boundaries. It is described by a fermionic Bogoliubov-de-Gennes (BdG) Hamiltonian featuring a particle-number-non-conserving pairing term that creates fermion pairs on nearest-neighbor sites. 

The bosonic analog of the Kitaev chain---a one-dimensional lattice model featuring nearest-neighbor bosonic pair creation---has recently garnered widespread attention as a minimal model for parametrically paired bosonic systems~\cite{Peano2016,Peano2016PRX,Mcdonald2018,qi2019bosonic,Ohashi2020,flynn2020deconstructing,Flynn2021,YokomizoMurakamiPRB2021,Chaudhary2021,wang2022quantum,Flynn_PRB2023,lv2024hidden,wang2025probing,bomantara2025floquet,Belyansky2025,bomantara2025nonhermitian,fortin2025topological,he2025hidden,lee2026symmetry,wang2026braiding}. 
Remarkably, while the underlying Hamiltonian is strictly Hermitian, the presence of parametric pairing alters the Heisenberg equations of motion, rendering the dynamical matrix (i.e., bosonic BdG Hamiltonian~\cite{ShindouPRB2013,LieuPRB2018,LeinSatoPRB2018,KawabataPRX2019,Ashida2020,Julku_PRL2021,Julku_PRB2021,OkumaPRB2022,Jalali_PRL2023,Iskin_PRA2023,OkumaPRB2024,Tesfaye_PRR2025,Deng_PRL2025,okuma2026steady,Yamamoto2026Ferro,tesfaye2026symplectic}) intrinsically non-Hermitian. 
Crucially, this form of non-Hermiticity requires no postselection, bypassing a major obstacle that typically hinders the experimental observation of non-Hermitian physics.
Recent experiments~\cite{Busnaina2024NatCommun,Slim2024} have demonstrated that the bosonic Kitaev chain can be realized in highly controllable setups, such as superconducting circuits~\cite{Busnaina2024NatCommun} and optomechanical devices~\cite{Slim2024}, successfully confirming predicted non-Hermitian physics.
However, despite such growing interest, Bose-Einstein condensation (BEC), a quintessential phenomenon of bosonic systems, has yet to be explored in this context.

In this work, we investigate the nonequilibrium BEC in an interacting bosonic Kitaev chain. Using mean-field theory, we demonstrate that BEC emerges when the parametric pairing strength exceeds a critical threshold. In the resulting BEC phase, the phase of the order parameter alternates between even and odd sites, giving rise to a spontaneous 
sublattice order.
The Bogoliubov excitations are effectively described by a non-Hermitian Dirac Hamiltonian, hosting
a re-entrant bosonic Andreev bound state (ABS) localized at the boundary.
At the critical point of the phase transition,
global exceptional points appear throughout the entire Brillouin zone. The resulting zero-energy excitations are equal-weight superpositions of quasiparticles and quasiholes that remain invariant under particle-hole transformations, realizing a bosonic analog of Majorana fermions
~\cite{Yamamoto2026Ferro}.





\begin{figure}[t]
    \centering
\includegraphics{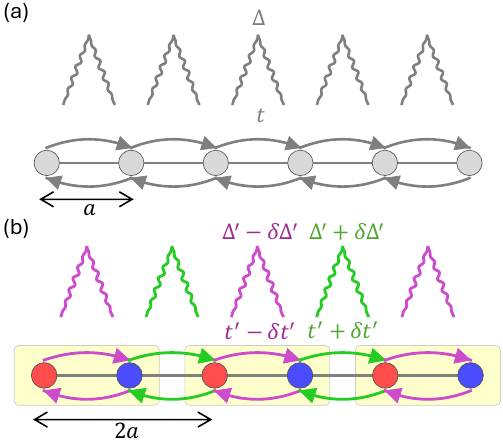}
    \caption{
    Schematics of (a) the bosonic Kitaev chain and (b)
    the Dirac BEC in the bosonic Kitaev chain. Emergent sublattices A and B are indicated in red and blue.
    }
    \label{figure:schematic}
\end{figure}

\textbf{\textit{Interacting bosonic Kitaev chain.}}---
We consider an interacting bosonic Kitaev chain with $N$ sites under open boundary condition.
The Hamiltonian is given by
\begin{align}
    H&=
    \sum_{i=0}^{N-2} it
    (c_i^\dagger c_{i+1}
    -c_{i+1}^\dagger c_i)
    +i\Delta 
    (c_i^\dagger c_{i+1}^\dagger
    -c_i c_{i+1})\notag\\ 
    &\hspace{3cm}+\sum_{i=0}^{N-1}
    \qty(
    2t c_i^\dagger c_i+\frac{U}{2}c_i^\dagger c_i c_i^\dagger c_i
    ),\label{eq:Hamiltonian of bosonic Kitaev chain}
\end{align}
which is schematically illustrated in Fig.~\ref{figure:schematic}(a).
Here, $c_i~(c_i^\dagger)$ is a bosonic annihilation (creation) operator for the site $i=0,1,2,...,N-1$, 
$t>0$ is a nearest-neighbor hopping amplitude,
$\Delta>0$ is a nearest-neighbor parametric pairing strength, and $U>0$ is an on-site repulsive interaction.
Crucially, the Hamiltonian remains Hermitian despite including pair creation and annihilation terms.
The Hamiltonian no longer possesses continuous global U(1) symmetry, as the parametric pairing term breaks particle-number conservation. However, it retains a discrete $\mathbb{Z}_2$ symmetry corresponding to a global $\pi$ phase shift of the field operator, i.e.$~c_i\rightarrow e^{i\pi}c_i$.

\textbf{\textit{Bose phase wave.}}---
We employ mean-field theory to examine the onset of BEC, defining the order parameter as the expectation value of the bosonic operator $\expval{c_i}=\sqrt{n_{i}}e^{i\phi_{i}}$~\cite{Shi1998},
where $n_i$ is the condensate density and $\phi_i$ is the local phase.
The mean-field energy is obtained as 
\begin{align}
    \expval{H}&=\sum_{i=0}^{N-2} 2\sqrt{n_i n_{i+1}}f_{i,i+1}
    +\sum_{i=0}^{N-1} \qty(
    2t n_i+\frac{U}{2} n_i^2
    ),\label{eq:mean-field-energy}
\end{align}
where the phase dependent part $f_{i,i+1}$ is given by
\begin{equation}
    f_{i,i+1}=t\sin(\phi_i-\phi_{i+1})
    +\Delta\sin(\phi_i+\phi_{i+1})\label{eq:phase dependent term}.
\end{equation}
At zero temperature, the ground-state configuration is obtained by minimizing Eq.~\eqref{eq:mean-field-energy}.
We employ an iterative numerical scheme initialized with random field configurations.

\begin{figure}[t]
    \centering
\includegraphics{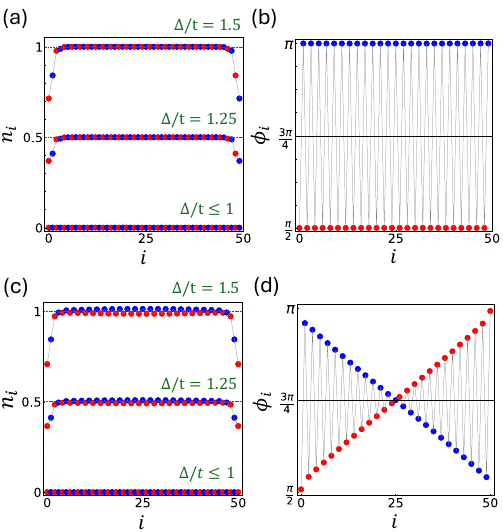}
    \caption{
    (a)(c) Condensate density $n_i$ for $\Delta/t\leq1$ and $\Delta/t=1.25,~1.5$, and (b)(d) local phase $\phi_i$ for $\Delta=1.5$. 
    System sizes are $N=50$ for (a)(b), and $N=51$ for (c)(d). Red and blue circles represent sites with even and odd indices, respectively. The dotted green horizontal line in (a) and (c) denotes the saturated bulk order parameter [Eq.~\eqref{eq:number of condensate boson in bulk}]. 
    Fixed parameter values are set to $t=1,~U=1$.
    }
    \label{figure:groud state}
\end{figure}

In Fig.~\ref{figure:groud state}(a), we plot the resulting site-resolved condensate density $n_i$ across representative values of $\Delta$, where we set the system size as $N=50$.
The red and blue dots represent the even and odd site indices, respectively.
We observe that when $\Delta<t$, the ground state of the system is just the bosonic vacuum $n_i=0$.
When $\Delta>t$, 
a finite BEC order parameter develops, signaling a nonequilibrium BEC transition driven by parametric pumping. 
Since the pairing term in the model [Eq.~\eqref{eq:Hamiltonian of bosonic Kitaev chain}] is induced by an external drive, the resulting 
BEC is intrinsically nonequilibrium, in contrast to equilibrium BECs in particle-conserving systems.
While the order parameter decays near the edge of the system, it approaches to the value 
\begin{align}
    \bar{n}=2(\Delta-t)/U, \label{eq:number of condensate boson in bulk}
\end{align}
in the bulk, which is depicted by a green dotted line in Fig.~\ref{figure:groud state}(a). 
In this condensation process, the repulsive interaction $U>0$ plays an indispensable role; without it, the system would collapse into an unstable state favoring an infinite number of particles.

In Fig.~\ref{figure:groud state}(b), we present the numerical results for the optimized phase $\phi_i$ of the order parameter at each site $i$, when the system is in a BEC phase ($\Delta>t$).
Interestingly, the system spontaneously shows a local phase modulation
\begin{align}
    \phi_{2m}=\frac{\pi}{2},~\phi_{2m+1}=\pi,\label{eq:ground state phase}
\end{align}
where $m$ is an arbitrary integer. The phase structure,
Eq.~\eqref{eq:ground state phase}, is unique up to a global $\pi$ phase shift, i.e.,$~\phi_i\rightarrow\phi_i+\pi$, which is consistent with the global $\mathbb{Z}_2$ symmetry of the original Hamiltonian [Eq.~\eqref{eq:Hamiltonian of bosonic Kitaev chain}].
This phenomenon can be understood as the bosonic analog of the pair density wave state in superconductivity~\cite{PDW_review_Agterberg,Yoshida2012,papaj2026pair}, where the superconducting order parameter oscillates periodically with the underlying crystalline lattice.
Thus, we generally refer to such phenomenon a Bose phase wave.
In the bosonic Kitaev model, we get an emergent two-sublattice system, labeled A (even sites) and B (odd sites), which is schematically illustrated in  Fig.~\ref{figure:schematic}(b). 

To understand why the Bose phase wave [Eq.~\eqref{eq:ground state phase}] is energetically favored, we focus on the phase-dependent terms [Eq.~\eqref{eq:phase dependent term}] in the mean-field energy [Eq.~\eqref{eq:mean-field-energy}]. 
Locally, $f_{i,i+1}$ is minimized by simultaneously satisfying conditions $\sin(\phi_i-\phi_{i+1})=-1$ and $\sin(\phi_i+\phi_{i+1})=-1$, which yields the unique optimal solution $(\phi_i,\phi_{i+1})=(\pi/2,\pi)$ up to a global $\pi$ phase shift.
While frustration makes it impossible to satisfy this configuration globally for $N>2$, numerical investigations show that the system compromises by forming these optimal dimers on every second bond.

As demonstrated above, when the total number of lattice sites $N$ is even, adjacent sites readily pair into dimers to achieve a globally stable configuration. 
Conversely, when $N$ is odd, complete dimerization is precluded and a single site remains unpaired.
Then, we performed the same analysis for $N=51$.
We plot the site-resolved condensate density and local phase for $N=51$ in Figs.~\ref{figure:groud state} (c)(d). 
Similar to the even-$N$ case, the system enters a BEC phase when $\Delta$ exceeds the same critical value, but exhibits key distinct features.
First, the condensate density spatially fluctuates around its bulk value [Eq.~\eqref{eq:number of condensate boson in bulk}].
Second, the phases on even and odd site indices twist globally along the chain in opposite directions between $\pi/2$ and $\pi$.
A detailed investigation of the odd-$N$ case lies beyond the scope of this work and is deferred to future study.
For the remainder of this paper, we focus exclusively on even lattice sites $N$.




\textbf{\textit{Bogoliubov excitation.}}---
To further investigate the physical properties of the Bose phase wave state, we examine its Bogoliubov excitations, which describe fluctuations around the BEC ground state.
The excitation spectrum is calculated by expanding the bosonic field operator as $c_i\rightarrow\expval{c_i}+\delta c_i$ and substituting it into the original Hamiltonian [Eq.~\eqref{eq:Hamiltonian of bosonic Kitaev chain}]. 
Expanding the Hamiltonian up to quadratic order in the fluctuations yields 
\begin{align}
    H-\expval{H}=\frac{1}{2}\psi^\dagger H_B\psi,
\end{align}
where $\psi:=(\delta c_0,~\delta c_1,\cdot\cdot\cdot,~\delta c_{N-1},~\delta c_0^\dagger,~\delta c_1^\dagger,\cdot\cdot\cdot,~\delta c_{N-1}^\dagger)^t$ is a $2N$-component bosonic Nambu spinor. The $2N\times2N$ Hermitian matrix $H_B$ is defined as 
\begin{equation}
    H_B=
    \mqty(
    h&s\\
    s^*&h^*
    ),
\end{equation}
with the normal part $h$ and the anomalous part $s$ given by $N\times N$ Hermitian and symmetric matrices, respectively.
The matrix elements are given by
\begin{align}
    h_{ii}&=2t+2Un_i,~h_{i,i+1}=h_{i+1,i}^*=it,\notag\\
    s_{ii}&=Un_ie^{2i\phi_i},~s_{i,i+1}=s_{i+1,i}=i\Delta,
\end{align}
with all other elements equal to zero.
For the normal part $h$, the on-site energy $h_{ii}$ is renormalized as $2t\rightarrow2t+2Un_i$
due to the interaction between the condensed bosons and the non-condensed bosons, supplementing the original nearest-neighbor hopping in Eq.~\eqref{eq:Hamiltonian of bosonic Kitaev chain}.
For the anomalous part $s$, in addition to the original nearest-neighbor pairing term in Eq.~\eqref{eq:Hamiltonian of bosonic Kitaev chain}, the on-site pairing term becomes finite, giving $s_{ii}=Un_i e^{2i\phi_i}=(-1)^{i+1}Un_i$, 
whose sign altenates between adjacent sites due to the effect of the intra-unit-cell bosonic phase modulation [Eq.~\eqref{eq:ground state phase}].

The dynamics of the Bogoliubov quasiparticles $\psi$ are governed by the Heisenberg equation of motion, 
yielding
$i\hbar\partial_t\psi=\comm{\psi}{H}=L\psi$, where we have used the bosonic commutation relation $\comm{\psi}{\psi^\dagger}=\sigma_3 \otimes 1_N$ in the second equality.
The $\sigma_i~(i=1,2,3)$ are the Pauli matrices acting on the particle-hole space, and $1_N$ is the $N\times N$ unit matrix.
Here $L$ is a $2N\times 2N$  matrix given by
\begin{equation}
    L=(\sigma_3 \otimes 1_N)H_B=
    \mqty(
    h&s\\
    -s^*&-h^*
    ),
\end{equation}
which is nothing but the bosonic BdG Hamiltonian.
A key feature is that when the pairing term $s$ is nonzero, $L$ becomes non-Hermitian.
Consequently, the energy spectrum  of the Bogoliubov excitations is obtained by solving the non-Hermitian eigenvalue problem:
\begin{align}
    L\ket{\Psi}=E\ket{\Psi}.\label{eq:schrodinger equation}
\end{align}

\begin{figure}[t]
    \centering
\includegraphics{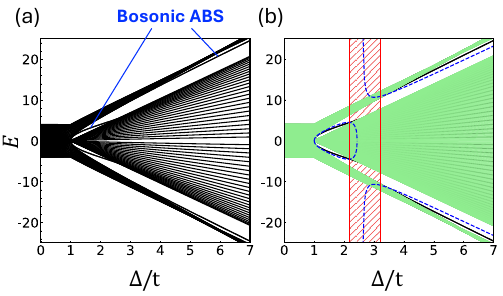}
    \caption{
    (a) Numerical Bogoliubov energy spectrum $E$ versus $\Delta$ for $N=50$ (black). (b) Analytical bulk (green) and edge (blue) energy spectra calculated from the effective non-Hermitian Dirac Hamiltonian. 
    The red shaded region indicates where $\lambda\leq a$.
    }
    \label{figure:bogoliubov sp}
\end{figure}

\begin{figure*}[t]
    \centering
\includegraphics{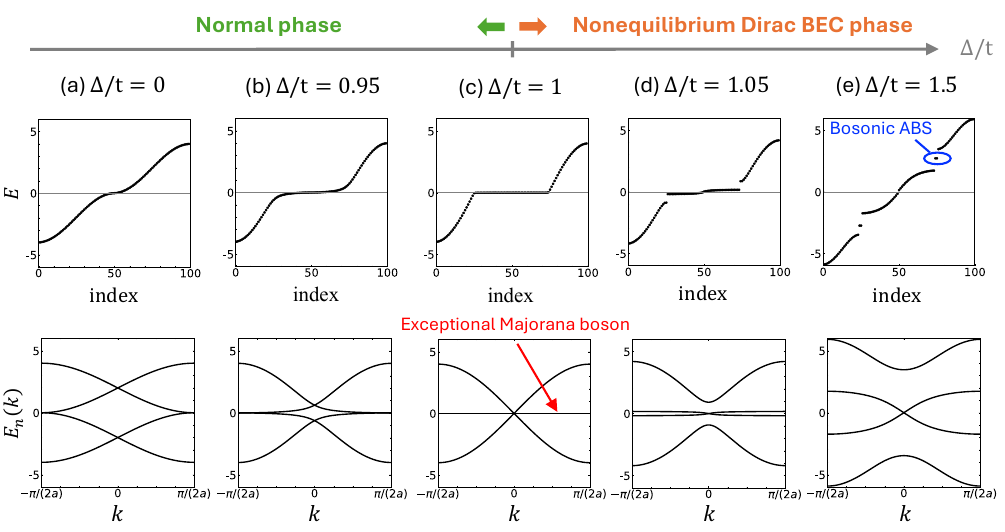}
    \caption{
    Upper panel: (a)–(e) Numerical Bogoliubov energy spectra $E$ for $N=50$ at representative $\Delta$ values. 
    Lower panel: (a)–(e) Analytical band structures $E_n(k)~(n=\pm 1,\pm 2)$ calculated from the effective non-Hermitian Dirac Hamiltonian at the same $\Delta$ values.
    }
    \label{figure:Dirac}
\end{figure*}

Figure~\ref{figure:bogoliubov sp}(a) shows the Bogoliubov energy spectrum $E$ as a function of $\Delta$ for system size $N=50$.
For clarity, the upper panels of Figs.~\ref{figure:Dirac}(a)-(e) plot the eigenvalues in ascending order for selected values of $\Delta$.
Within the normal phase $(0\leq\Delta<t)$, the energy eigenvalues cluster around zero energy as $\Delta$ increases.
At the critical point $\Delta=t$, exactly $N$ out of the $2N$ total eigenvalues vanish [Fig.~\ref{figure:Dirac}(c)], reflecting anomalously enhanced fluctuations due to the phase transition.
Upon entering the BEC phase $(\Delta>t)$, all zero-energy states disappear, marking a transition to a thermodynamically stable phase, and energy gaps open in both the particle and hole sectors. 
Crucially, two degenerate bosonic Andreev bound states (ABS) localized at the boundary emerge within each energy gap, as highlighted in Fig.~\ref{figure:bogoliubov sp}(a).
These bound states exhibit unusual reentrant behavior:
with increasing $\Delta$, they emerge from the upper energy branch, move downward to be absorbed by the lower branch, and finally reappear from the upper branch as $\Delta$ increases further.

\textbf{\textit{Effective model.}}---
To gain physical insight and qualitatively understand the Bogoliubov excitation from the previous section, we construct an effective model to describe it.
To this end, we neglect the decay of the order parameter near the system boundaries [Fig.~\ref{figure:groud state}(a)] and assume a uniform condensate density across all sites:
$n_i=\bar{n}$~[Eq.~\eqref{eq:number of condensate boson in bulk}] in the BEC phase and $n_i=0$ in the normal phase.

Next,
we apply a purely local Bogoliubov transformation~\cite{Peano2016} that only mixes bosonic particles and holes residing on the same site. In this new basis, the $s_{ii}$-component is eliminated, simplifying the effective Hamiltonian to
\begin{align}
    L_\mathrm{eff}=
    \mqty(
    h_\mathrm{eff}&s_\mathrm{eff}\\
    -s^{*}_\mathrm{eff}&-h^{*}_\mathrm{eff}
    ),\label{eq:L_eff}
\end{align}
where the matrix elements of $h_\mathrm{eff}$ and $s_\mathrm{eff}$ are given by
\begin{align}
    (h_\mathrm{eff})_{ii}&=\bar{E},~(h_\mathrm{eff})_{i,i+1}=(h_\mathrm{eff})_{i+1,i}^*=i(t'+(-1)^{i+1}\delta t'),\notag\\
    (s_\mathrm{eff})_{ii}&=0,~(s_\mathrm{eff})_{i,i+1}=(s_\mathrm{eff})_{i+1,i}=i(\Delta'+(-1)^{i+1}\delta\Delta'),
\end{align}
with all other elements equal to zero.
Here, the on-site energy is given by $\bar{E}=\sqrt{(2t+2U\bar{n})^2-(U\bar{n})^2}$, and the hopping amplitude and pairing strength are written by using the following quantities; 
\begin{align}
    t'&=t\cosh\xi,~\delta t'=\Delta\sinh\xi,\notag\\
    \Delta'&=\Delta\cosh\xi,~\delta\Delta'=t\sinh\xi,
\end{align}
where $\cosh \xi=(2t+2U\bar{n})/\bar{E}$, and $\sinh\xi=U\bar{n}/\bar{E}$.
As illustrated in Fig.~\ref{figure:schematic}(b),
the effective intra- and inter-unit-cell hopping amplitudes are given by $t'\pm\delta t'$, respectively.
In the same way, the effective intra- and inter-unit-cell pairing strength are given by $\Delta'\pm\delta \Delta'$, respectively.

\textbf{\textit{Non-Hermitian Dirac Hamiltonian.}}---
As a first step, we examine the bulk eigenstates of the effective model, which extend throughout the entire system. Because the unit cell spans twice the lattice constant $a$, its position is indexed by
$x_j=2aj$,
and the corresponding Brillouin zone is reduced to $-\pi/(2a)\leq k\leq\pi/(2a)$.
We adopt a Bloch ansatz, $\ket{\Psi(x_j)}=e^{ikx_j} \ket{u_{nk}}$, where the periodic part $\ket{u_{nk}}$ is a four-component vector accounting for both particle–hole and sublattice degrees of freedom. 
Substituting this ansatz into the Schrodinger equation [Eq.~\eqref{eq:schrodinger equation}] yields the reduced form
\begin{align}
    L_\mathrm{eff}(k)\ket{u_{nk}}=E_n(k)\ket{u_{nk}},    
\end{align}
where $L_\mathrm{eff}(k)$ is a Bloch Hamiltonian and
can be cast directly into the form of a non-Hermitian Dirac Hamiltonian:
\begin{align}
    L_\mathrm{eff}(k)=\bar{E}\alpha_4+a_k\alpha_{13}+b_k\alpha_{32}+c_ki\alpha_1+d_k i\alpha_2,\label{eq:L_eff(k)}
\end{align}
where $a_k=-(t'-\delta t')+(t'+\delta t')\cos k,~b_k=-(t'+\delta t')\sin k,~c_k=(\Delta'-\delta\Delta')+(\Delta'+\delta\Delta')\cos k,~d_k=(\Delta'+\delta\Delta')\sin k$ are real coefficients.
The Hermitian matrix $\alpha_\mu~(\mu=1,2,3,4,5)$ denotes the standard Dirac $\alpha$-matrices, commonly used in the construction of the Dirac Hamiltonian.
Their explicit forms are given by $\alpha_i=\sigma_x\otimes\sigma_i(i=1,2,3),~\alpha_4=\sigma_z\otimes1,~\alpha_5=\sigma_y\otimes1$, and they satisfy the Clifford algebra anti-commutation relations $\pb{\alpha_\mu}{\alpha_\nu}=2\delta_{\mu\nu}$.
Together with the additional Hermitian matrices $\alpha_{\mu\nu}=i\alpha_\mu\alpha_\nu~(\mu\ne\nu)$ and the unit matrix, these 16 matrices constitute a complete basis for the real vector space of $4\times 4$ Hermitian matrices.
Because the last two terms in Eq.~\eqref{eq:L_eff(k)} are proportional to $i\alpha_1$ and $i\alpha_2$, the Bloch Hamiltonian becomes non-Hermitian.

The system consists of 4 bands, denoted $E_n(k)$ for $n=\pm1,\pm2$.
Positive indices correspond to particle bands and negative indices to hole bands, ordered such that $E_{-2}(k)\leq E_{-1}(k)\leq E_1(k)\leq E_2(k)$.
Owing to the simplicity of Eq.~\eqref{eq:L_eff(k)}, its eigenvalues $E_n(k)$ can be obtained analytically as 
\begin{align}
    E_n(k)&=\pm\sqrt{\bar{E}^2+a_k^2+b_k^2-c_k^2-d_k^2\pm2M_k},\label{eq:energy band}
\end{align}
where $M_k=\sqrt{\bar{E}^2(a_k^2+b_k^2)-(a_k d_k+b_k c_k)^2}$.

In Fig.~\ref{figure:bogoliubov sp}(b), this analytical result (green) is plotted as a function of $\Delta$, overlaying the numerical results (black). 
We observe that the analytic spectrum $E_n(k)$ is in perfect agreement with the numerical bulk spectrum.
The lower panel of Fig.~\ref{figure:Dirac}(a)-(e) shows the band structure $E_n(k)$ as a function of $k$ for representative values of $\Delta$.
In the normal phase $(0\leq\Delta<t)$, increasing $\Delta$ causes the two bands $E_{\pm1}(k)$ to move closer together. 
Notably, because the A and B sublattices are equivalent in the noramal phase, the true spatial periodicity is $a$ rather than $2a$.
Consequently, our chosen Brillouin zone is halved relative to the true first Brillouin zone, resulting in artificial band folding that doubles the number of bands.
For instance, at $\Delta=0$, $E_1(k)$ and $E_2(k)$ reduce to a single dispersion, $2t(1-\sin ka)$.
At the critical point $(\Delta=t)$, the particle and hole bands $E_{\pm1}(k)$ merge into a completely flat, zero-energy band, giving rise to an $N$-fold degeneracy. 
Rather than a simple degeneracy, this represents a global exceptional point characterized by the simultaneous coalescence of eigenvalues and eigenvectors
throughout the entire Brillouin zone.
The corresponding eigenstate is given by $\ket{u_{1k}}=\ket{u_{-1k}}\propto(-i,1,i,1)^t$,
which is invariant under particle-hole symmetry $C=(\sigma_x\otimes 1)K$, where $K$ is a complex conjugate operator.
Thus we call such a state an exceptional Majorana boson~\cite{Yamamoto2026Ferro}.
Finally, in the BEC phase $(\Delta>t)$, the equivalence between A and B sites is broken; the two-sublattice structure becomes physical, and an energy gap opens at $k=0$.

Next, we discuss the boundary states.
We demonstrate that the bosonic Andreev bound state localized at the boundary can be understood via the evanescent modes of the effective non-Hermitian Dirac Hamiltonian. 
Specifically, the left edge evanescent mode is characterized by an imaginary wave vector $k=i\lambda^{-1}$, 
and its eigenstate takes the form
\begin{align}
    \ket{\Psi_\mathrm{edge}(x_j)}\propto e^{-x_j/\lambda}\mqty(
    i(E_\mathrm{edge}+\bar{E})\\
    \frac{2\Delta t}{(t'+\delta t')}\\
    -i(E_\mathrm{edge}-\bar{E})\\
    -\frac{2\Delta t}{(t'+\delta t')}
    )
\end{align}
with the localization length $\frac{1}{\lambda}=\frac{1}{2a}\log \qty|\frac{t'+\delta t'}{t'-\delta t'}|$,
and the eigenenergy
\begin{align}
    E_\mathrm{edge}=\sqrt{\bar{E}^2-\frac{(2t\Delta)^2}{t'^2-\delta t'^2}e^{-2\xi}}.\label{eq:edge energy}
\end{align}
The right edge state can be obtained in an analogous manner. These two edge states are degenerate in energy owing to the spatial inversion symmetry of the system.

In Fig.~\ref{figure:bogoliubov sp}(b),
this analytical edge state energy [Eq.~\eqref{eq:edge energy}] is shown in a blue dotted curve.
The red shaded region highlights where the localization length $\lambda$ is shorter than the lattice constant $a$, rendering the effective model invalid. Within the valid regime (outside the red region), the model agrees qualitatively with numerical results and accurately captures the re-entrant behavior of the bosonic Andreev bound states.


\textbf{\textit{Conclusion.}}---
We demonstrate the emergence of a non-equilibrium BEC in an interacting bosonic Kitaev chain.
Increasing the parametiric pairng strength above the critical value, we discover the superfluid order with the intra-unit-cell bosonic phase modulation, which can be interpreted as a bosonic analog of the pair density wave in fermionic superconductivity~\cite{PDW_review_Agterberg,Yoshida2012,papaj2026pair}.
The elementary Bogoliubov excitations are naturally captured by a non-Hermitian Dirac Hamiltonian, which displays novel features, including an exceptional-point-induced Majorana bosons at the critical point of the phase transition, and re-entrant bosonic Andreev bound states localized at the system edges.
Our model can be readily implemented in artificial experimental platforms, including superconducting circuits and optomechanical systems~\cite{Busnaina2024NatCommun,Slim2024}.
For potential applications,
our theory
can be applied to nonequilibrium condensates, such as magnon BECs~\cite{Demokritov2006,Bozhko2016,Bozhko2019,Divinskiy2021,Yamamoto2025,Yamamoto2026Ferro,bailey2026multiband} and exciton-polariton BECs~\cite{Byrnes2014,lerario2017room,hanai2019non,hanai2020critical,Morimoto2020}.

\textbf{\textit{Acknowledgment.}}---
This work was supported in part by JSPS KAKENHI Grants No. JP22H04933, JP23K17353,
JP24K21530, JP24H00007, JP25H01249, JP26H02016, Japan.
\bibliography{reference}
\clearpage
\begin{widetext}
\appendix

\end{widetext}


\end{document}